\RequirePackage{fix-cm}
\documentclass[twocolumn,epjc3,showpacs,preprintnumbers,amsmath,amssymb]{svjour3}
\smartqed  
\RequirePackage{graphicx}
\RequirePackage[numbers,sort&compress]{natbib}
\usepackage{graphicx,amsmath,amsfonts,latexsym,amssymb,graphics,epsfig,subfigure,color,makeidx,hyperref}
\usepackage{multirow}

\journalname{Eur. Phys. J. C}

\newcommand\beq{\begin{equation}}
\newcommand\eeq{\end{equation}}
\newcommand\beqn{\begin{eqnarray}}
\newcommand\eeqn{\end{eqnarray}}
\newcommand\bal{\begin{align}}
\newcommand\eal{\end{align}}
\newcommand\nn{\nonumber}
\newcommand\fc{\frac}
\newcommand\lt{\left}
\newcommand\rt{\right}
\newcommand\pt{\partial}

\begin{document}

\allowdisplaybreaks

\title{Tensor stability in Born-Infeld determinantal gravity}

\author{Ke Yang\thanksref{e1,addr1}
       \and Yu-Peng Zhang\thanksref{e2,addr2}
       \and  Yu-Xiao Liu\thanksref{e3,addr2,addr3}
}

\thankstext{e1}{e-mail: keyang@swu.edu.cn}
\thankstext{e2}{e-mail: zhangyupeng14@lzu.edu.cn}
\thankstext{e3}{e-mail: liuyx@lzu.edu.cn, corresponding author}

\institute{
School of Physical Science and Technology, Southwest University, Chongqing 400715, China \label{addr1}
\and
Institute of Theoretical Physics $\&$ Research Center of Gravitation, Lanzhou University, Lanzhou 730000, China\label{addr2}
\and
Key Laboratory for Magnetism and Magnetic of the Ministry of Education, Lanzhou University, Lanzhou 730000, China\label{addr3}
 }

\maketitle

\begin{abstract}

We consider the transverse-traceless tensor perturbation of a spatial flat homogeneous and isotropic spacetime in Born-Infeld determinantal gravity, and investigate the evolution of the tensor mode for two solutions in the early universe. For the first solution where the initial singularity is replaced by a regular geometric de Sitter inflation of infinite duration, the evolution of the tensor mode is stable for the parameter spaces $\alpha<-1$, $\omega\geq-1/3$ and $\alpha=-1$, $\omega>0$. For the second solution where the initial singularity is replaced by a primordial brusque bounce, which suffers a sudden singularity at the bouncing point, the evolution of the tensor mode is stable for all regions of the parameter space. Our calculation suggests that the tensor evolution can hold stability in large parameter spaces, which is a remarkable property of Born-Infeld determinantal gravity. We also constrain the theoretical parameter $|\lambda|\geq 10^{-38} \text{m}^{-2}$ by resorting to the current bound on the speed of the gravitational waves.
\end{abstract}

\maketitle

\section{Introduction}
The teleparallel equivalent of general relativity, also called  teleparallel gravity or teleparallelism for short, can be traced back to an attempt by Einstein to unify the electromagnetism and gravity on the mathematical structure of distant parallelism \cite{Einstein1928}. In this theory, there is a set of dynamical vierbein (or tetrad) fields which form the orthogonal bases for the tangent space of each spacetime point. Instead of the Levi-Civita connection $\{{^{\rho}}_{\mu\nu}\}$, a spacetime is characterized by a curvature-free Weitzenb$\ddot{\text{o}}$ck connection ${\Gamma^\rho}_{\mu\nu}={e_{A}}^{\rho}\pt_{\nu}{e^{A}}_{\mu}$, with ${e^{A}}_{\mu}$ the vierbein, which refers to the metric through the relation $g_{\mu\nu}={e^{A}}_{\mu}{e^{B}}_{\nu}\eta_{AB}$, where $\eta_{AB} = \text{diag}(1, -1, -1, -1)$ is the Minkowski metric for the tangent space. Although the Weitzenb$\ddot{\text{o}}$ck spacetime is flat, it possesses torsion, which is defined as ${T^{\rho}}_{\mu\nu}={\Gamma^{\rho}}_{\nu\mu}-{\Gamma^{\rho}}_{\mu\nu}$.

The well-known action of teleparallel gravity reads
\beq
S_{\text{TG}}=\fc{1}{16\pi G}\int{d^{d+1}x\, e\,T},
\label{Action_teleparallel}
\eeq
where $e=|{e^A}_M|=\sqrt{-|g_{\mu\nu}|}$, and $T\equiv{S_\rho}^{\mu\nu}{T^{\rho}}_{\mu\nu}$ is torsion scalar contracted by the torsion tensor and a new tensor ${S_\rho}^{\mu\nu}$ defined by ${S_\rho}^{\mu\nu}=\fc{1}{2}({K^{\mu\nu}}_{\rho}+\delta^{\mu}_{\rho}{T^{\sigma\nu}}_{\sigma}-\delta^{\nu}_{\rho}{T^{\sigma\mu}}_{\sigma})$, with ${K^{\mu\nu}}_{\rho}$ the contorsion tensor related to the difference between the Levi-Civita connection and
Weitzenb$\ddot{\text{o}}$ck connection, i.e., ${K^\rho}_{\mu\nu}=\Gamma^{\rho}_{\mu\nu}-\{{^{\rho}}_{\mu\nu}\}
=\fc{1}{2}({{T_\mu}^{\rho}}_{\nu}+{{T_\nu}^{\rho}}_\mu-{{T^\rho}_{\mu}}_{\nu})$. Gravitational interaction is described by the curved spacetime geometry in general relativity, however, the contorsion tensor can be regarded as a gravitational
force acting on particles in teleparallel gravity.  Nevertheless, no matter which description of  gravity we use, with the identity $T=-R+2 e^{-1} \pt_\nu (e T_{\mu}{}^{\mu\nu})$ between the torsion scalar and Ricci scalar, the equivalence between teleparallel gravity and general relativity is manifest from the action (\ref{Action_teleparallel}).

As the cornerstone of modern cosmology, general relativity provides precise descriptions to a variety of phenomena in our universe. However, it is well-known that general relativity suffers from various troublesome theoretical problems, such as the dark matter problem \cite{Bertone2005}, dark energy problem \cite{Li2011a} and the unavoidable singularity problem \cite{Hawking1970}. One of the attempts to solve the singularity problem in classical level, first suggested by Deser and Gibbons \cite{Deser1998}, follows the spirit of Born-Infeld electromagnetic theory, which regularises the divergent self-energy of the electron in classical dynamics \cite{Born1934}. One of the Born-Infeld type generalized gravity can be written as the form \cite{Vollick2004,Banados2010,Ferraro2007,Ferraro2010,Fiorini2013,Chen2016,Chen2017a}
\beq
S\!=\!\frac{\lambda}{16\pi G}\int{}d^{d+1}x\lt[\sqrt{-|g_{\mu\nu}\!+\!{2}{\lambda}^{-1} F_{\mu\nu}|}\!-\!\Delta\sqrt{-|g_{\mu\nu}|} \rt],
\eeq
where the rank-2 tensor $F_{\mu\nu}$ is a function of certain fields $\psi_i$ and their derivatives, $\lambda$ is the Born-Infeld constant with mass dimension 2, and $\Delta$ is a constant.

In low-energy limit  ($\lambda\rightarrow\infty$), the above action approximates to
\beq
S\approx\frac{1}{16\pi G}\int{}d^{d+1}x\sqrt{-|g_{\mu\nu}|}\lt[\text{Tr}(F_{\mu\nu})+2\Lambda\rt],
\eeq
where $\Lambda\equiv(1-\Delta)\lambda/2$ is the cosmological constant.

In order to recover a proper low-energy theory, the simplest case is to choose $F_{\mu\nu}$ to be the Ricci tensor $R_{\mu\nu}(\psi_i)$, then general relativity is recovered. However, If $\psi_i$ is the metric field, namely working in a pure metric formalism, it will lead to fourth order field equations with ghost instabilities \cite{Deser1998}. If $\psi_i$ is the connection field, namely working in the Palatini formalism, the theory is free from the ghost problem \cite{Vollick2004,Banados2010}, and is now dubbed Eddington-inspired Born-Infeld (EiBI) gravity. An intriguing property of EiBI theory is that it may avoid the initial Big Bang singularity of the universe \cite{Banados2010,Cho2012,Avelino2012a}. However, although the background evolution maybe free from an initial singularity, by including the linear perturbations, the overall evolution may still be singular \cite{Escamilla-Rivera2012,Yang2013}.

Another interesting choice is to simply require $\text{Tr}(F_{\mu\nu})$ to be the torsion scalar \cite{Ferraro2007,Ferraro2010,Fiorini2013}. Thus, the low-energy theory recovers the teleparallel gravity or general relativity equivalently. This can be fulfilled with $F_{\mu\nu}=\alpha F^{(1)}_{\mu\nu}+\beta F^{(2)}_{\mu\nu}+\gamma F^{(3)}_{\mu\nu}$ with $F^{(1)}_{\mu\nu}={S_\mu}^{\rho\sigma}T_{\nu\rho\sigma}$, $F^{(2)}_{\mu\nu}={S_{\rho\mu}}^{\sigma}{T^{\rho}}_{\nu\sigma}$,  $F^{(3)}_{\mu\nu}=g_{\mu\nu}T$, and $\alpha+\beta+(d+1)\gamma=1$. This theory leads to  second-order field equations, and is dubbed Born-Infeld determinantal gravity. It supports some cosmological solutions by replacing the possible initial singularity with a de-Sitter phase or a bounce \cite{Fiorini2013,Fiorini2016}. The authors in Ref. \cite{Bouhmadi-Lopez2014a} pointed out that although the theory is singularity-free in some regions of the parameter space, nevertheless, the Big Rip, Big Freeze, or Sudden singularities may still emerge in some other regions of the parameter space. The equations of motion were analyzed and Schwarzschild geometry was studied in Ref. \cite{Fiorini2016a}. If $\alpha=\beta=0$, the theory reduces to an $f(T)$ type theory, a spatially flat cosmology and a 5-dimensional domain wall have been considered in this reduced Born-Infeld-$f(T)$ theory \cite{Jana2014,Yang2018}.

The property of tensor perturbations has been widely analyzed in torsional theories, such as $f(T)$, $f(T, B)$, $f(T, T_G)$ and extended symmetric teleparallel gravities, see for instance \cite{Chen2011,Wu2012a,Izumi2013,Cai2018,Li2018,Farrugia2018,Soudi2018}. In this work, we investigate the evolution of transverse-traceless (TT) tensor perturbation in early high energy regime of the spatially flat Friedmann-Robertson-Walker (FRW) cosmology in Born-Infeld determinantal gravity. The evolution of TT tensor mode, which is pure gravitational and is irrelevant to matter density perturbations, reveals the overall stability of the singularity-free background solutions against the tensor perturbation.  tensor perturbation analysis  in torsional theories Through the paper, the capital Latin indices $A, B, \cdots$ and small Latin indices $a, b, \cdots$ label the four-dimensional and  three-dimensional coordinates of tangent 	space, respectively, and Greek indices $\mu,\nu, \cdots$ and small Latin indices $i, j, \cdots$ label the four-dimensional spacetime and three-dimensional space coordinate, respectively. For simplicity, we set a vanishing cosmological constant by fixing $\Delta=1$ in the following analysis.

The paper is organized as follows. In Sec. \ref{Sec_EoM}, we introduce the equations of motion of Born-Infeld determinantal gravity. In Sec. \ref{Sec_Perturbation}, we consider the tensor perturbation in a spatial flat FRW cosmological background and get the evolution of tensor mode. In Sec. \ref{Sec_Stability}, we investigate the evolution of tensor mode in very early cosmology for two specific cosmic solutions. In Sec. \ref{Sec_Parameter_Constraint}, we constrain the theoretical parameter $\lambda$ with the current bound on the speed of the gravitational waves. Finally, brief conclusions are presented.

\section{Equations of motion}\label{Sec_EoM}

We start from the action in which gravity is minimally coupled to a matter field \cite{Fiorini2013}
\begin{align}
S&=\frac{\lambda}{16\pi G}\int{}d^{d+1}x\lt[\sqrt{-|g_{\mu\nu}+2\lambda^{-1} F_{\mu\nu}|}-\sqrt{-|g_{\mu\nu}|} \rt]\nn\\
&+\int{}d^{d+1}x \mathcal{L}_M,
\end{align}
where $\mathcal{L}_M$ represents the Lagrangian of a matter field coupling only to the vierbein field or to metric equivalently. By varying with respect to the vierbein, one gets the Euler-Lagrange equation
\beq
\frac{\pt \mathcal{L}_G}{\pt e^A{}_\mu}-\pt_\gamma\lt(\frac{\pt \mathcal{L}_G}{\pt (\pt_\gamma e^A{}_\mu)}\rt)=\frac{16 \pi  G e}{\lambda }{\Theta_A}^\mu,
\eeq
where $\mathcal{L}_G$ represents the gravitational Lagrangian,
and ${\Theta_A}^\mu\equiv-e^{-1}{\pt  \mathcal{L}_M}/{\pt e^A{}_\mu}$ is related to the energy-momentum tensor ${\Theta}^{\mu\nu}={e^A}_\mu {\Theta_A}^\nu$. If the action of the matter field is local Lorentz invariant, then the energy-momentum tensor is symmetric and conserved \cite{Li2011d}.

With some algebra, the left two terms in Euler-Lagrange equation can be written explicitly as
\begin{align}
&\frac{\pt \mathcal{L}_G}{\pt e^A{}_\mu}=\frac{|U_{\mu\nu}|^{\frac{1}{2}}(U^{-1})^{\beta\alpha} }{2} \lt(e_{A(\alpha} \delta^\mu{}_{\beta)}+\frac{2 \pt F_{\alpha\beta}}{\lambda  \pt e^A{}_\mu}\rt)-{e_A}^\mu e,\\
&\frac{\partial L_G}{\pt (\pt_\gamma e^A{}_\mu)}=\frac{|U_{\mu\nu}|^{\frac{1}{2}}(U^{-1})^{\beta\alpha}}{\lambda}\frac{ \pt F_{\alpha\beta} }{\pt (\pt_\gamma e^A{}_\mu)},
\end{align}
where $U_{\mu\nu}=g_{\mu\nu}+2\lambda^{-1} F_{\mu\nu}$.
After contracting the index $A$ of tangent space via multiplying a vierbein ${e^A}_\nu$, the equations of motion read \cite{Fiorini2016a}
\begin{align}
&\frac{|U_{\mu\nu}|^{\frac{1}{2}} \lt({U}^{-1}\rt)^{\beta\alpha}  }{2}\lt[{\delta^\mu}_{(\alpha} g_{\nu\beta)} + \frac{2e^A{}_{\nu}}{\lambda} \frac{\partial F_{\alpha\beta}}{ \partial {e^A}_\mu} \rt]-{\delta^\mu}_\nu e\nn\\
&-\frac{e^A{}_{\nu}}{\lambda }\partial_\gamma\lt[ |U_{\mu\nu}|^{\frac{1}{2}}  \lt({U}^{-1}\rt)^{\beta\alpha} \frac{\partial F_{\alpha\beta}}{\partial(\partial_\gamma {e^A}_\mu)} \rt]
=\frac{16 \pi  G e}{\lambda } {\Theta_\nu}^\mu,
\label{EoM}
\end{align}
where the energy-momentum tensor for a perfect fluid reads $\Theta ^\mu{}_\nu=(\rho+P) u^\mu u_\nu-P \delta ^\mu{}_\nu$, and the two partial  derivative terms are written explicitly as
\begin{align}
&\frac{\partial F_{\alpha\beta}}{ \partial {e^A}_\mu}=\alpha
\lt({\delta^\mu}_\alpha F^{(1)}_{A\beta}+{\delta^{\mu}}_{\beta} F^{(1)}_{\alpha A} +{Q^\mu}_{A\alpha\rho\sigma} {T_\beta}^{\rho\sigma}\nn\rt.\\
&\lt.-2S_{\alpha\rho(A}{T_\beta}^{\rho\mu)}\rt)
+\beta \lt({Q^\mu}_{A\rho\alpha\sigma} {{T^\rho}_{\beta}}^\sigma - {S_{\rho\alpha}}^{(\mu}{T^\rho}_{\beta A)} \rt)\nn\\
&+\gamma \lt({\delta^\mu}_{(\alpha} e_{A\beta)}T-4g_{\alpha\beta}{F^{(2)\mu}}_{A}\rt),\\
&\frac{\partial F_{\alpha\beta}}{\partial(\partial_\gamma {e^A}_\mu)}=\alpha\lt(2e_{A\beta} {S_\alpha}^{\gamma\mu}+ {D_{\alpha\rho\sigma A}}^{[\gamma\mu]} {T_\beta}^{\rho\sigma} \rt)\nn\\
&+\beta\lt({S_{A\alpha}}^{[\mu} {\delta ^{\gamma]}}_\beta + {D_{\rho\alpha\sigma A}}^{[\gamma\mu]} {{T^{\rho}}_\beta}^\sigma \rt)+4 \gamma g_{\alpha\beta} {S_{A}}^{\gamma\mu},
\end{align}
with the tensors $ {S^C}_{\alpha\beta}$ and ${{{Q^\lambda}_A}^C}_{\alpha\beta}$ defined as $ {S^C}_{\alpha\beta}={{D^C}_{\alpha\beta B}}^{\rho\sigma}{T^B}_{\rho\sigma}$ and ${{{Q^\lambda}_A}^C}_{\alpha\beta}=\frac{\pt {S^C}_{\alpha\beta}}{\pt {e^A}_\lambda}$, and given by
\begin{align}
{{D^C}_{\alpha\beta B}}^{\rho\sigma}&=\frac{1}{4}\lt({\delta_\alpha}^\rho {\delta_\beta}^\sigma {\delta_B}^C-e^{C\sigma}e_{B[\alpha}{\delta_{\beta]}}^\rho \rt)\nn\\
&+\frac{1}{2}{e_B}^\sigma{e^C}_{[\alpha}\delta_{\beta]}^\rho,\\
{{{Q^\lambda}_A}^C}_{\alpha\beta}&=\frac{1}{4}\lt(e^{C\lambda}T_{[\alpha\beta]A}-{\delta^\lambda}_{[\alpha}{T_{A\beta]}}^C\rt)\nn\\
&-\frac{1}{2}\lt({\delta_A}^C {\delta^\lambda} _{[\alpha}
{T^\sigma}_{\sigma\beta]}-{e^C}_{[\alpha}{T^\lambda}_{A\beta]}\right).
\end{align}

\section{Linear tensor perturbation}\label{Sec_Perturbation}

We consider a 4-dimensional perturbed spatial flat FRW spacetime ($d=3$) with the metric to be of the form
\beq
ds^2=dt^2-a^2(t)\lt[ \delta_{ij}+2 h_{ij}(t,x)  \rt] dx^i dx^j,
\eeq
where $a(t)$ is the scale factor and $h_{ij}(t,x)$ a TT tensor perturbation, i.e.,
$\pt^i h_{ij}=\delta^{ij} h_{ij}=0$. The corresponding perturbed vierbein reads
\beqn
{e^A}_\mu=\left( {\begin{array}{*{20}{c}}
1&  0\\
0  & a(t)\left({\delta^{a}}_i+{h^{a}}_i \right)
\end{array}} \right),
\eeqn
where ${h^{a}}_i={\delta^{a}}_j {h^{j}}_i$. With the perturbed vierbein, the nonvanishing components of  perturbed torsion tensor are
\beqn
{T^{k }}_{0 j }&=&-{T^{k }}_{j 0}=H {\delta^{k }}_{j }+\dot{h}^{k}{}_{j },\\
{T^{k }}_{ij }&=&\partial _{[i }h^{k }{}_{j ]},
\eeqn
where $H=\dot{a}/a$ is the Hubble parameter. Then, the nonvanishing components of the perturbed contorsion tensor read
\beqn
{K^{k }}_{0 j }&=&-H {\delta ^{k }}_{j}-\dot{h}^{k }{}_{j},\\
{K^0}_{ij}&=&-a^2 H \delta _{ij }+2 H h_{ij }+\dot{h}_{ij},\\
{K^{k }}_{ij }&=&\partial ^{k }h_{ij }-\pt_i{h^{k }}_j,
\eeqn
and the nonvanishing components of the perturbed tensor ${S_P}^{MN}$ read
\beqn
{S_{k }}^{i 0}&=&\frac{1}{2} \lt[(d-1) H \delta ^{i }{}_{k }-\dot{h}^{i }{}_{k }\rt],\\
{S_{k }}^{0 j }&=&-\frac{1}{2} \lt[(d-1) H \delta ^{j}{}_{k}-\dot{h}^{j }{}_{k }\rt],\\
{S_{k }}^{ij }&=&-\frac{1}{2}a^{-2}\pt ^{[i }h^{j ]}{}_{k }.
\eeqn

The perturbation of $F_{\mu\nu}$ can be assembled by  $F_{\mu\nu}=\alpha F^{(1)}_{\mu\nu}+\beta F^{(2)}_{\mu\nu}+\gamma F^{(3)}_{\mu\nu}$, with the nonvanishing components of $F^{(1)}_{\mu\nu}$, $F^{(2)}_{\mu\nu}$, and $F^{(3)}_{\mu\nu}$ given as
\begin{align}
F^{(1)}_{ij}&=(d-1)a^2 H^2 \lt(\delta _{ij }+2 h_{ij }+\frac{d-2}{d-1} \frac{ \dot{h}_{ij }}{H}\rt),\\
F^{(2)}_{00}&=-\frac{1}{2} d (d-1) H^2,\\
F^{(2)}_{ij }&=\frac{d-1}{2} a^2 H^2\lt(\delta _{ij }+2 h_{ij }+\frac{d-2}{d-1} \frac{\dot{h}_{ij }}{H}\rt),\\
F^{(3)}_{00}&=T=-d (d-1) H^2,\\
F^{(3)}_{ij }&=d (d-1) a^2 H^2 \lt(\delta _{ij}+2 h_{ij }\rt).
\end{align}
The expressions for the perturbations of ${\partial F_{\alpha\beta}}/{ \partial {e^A}_\mu}$ and ${\partial F_{\alpha\beta}}/{\partial(\partial_\gamma {e^A}_\mu)}$ are listed in appendix \ref{appendix}.

By only focusing on the TT tensor mode, we can shut down all the scalar and vector modes in the perturbed perfect fluid, since the scalar, vector and tensor modes are decoupled from each other and evolve separately. So the nonvanishing components of the perturbed energy-momentum tensor are simply given by
\beq
\Theta ^0{}_0=\rho ,  \quad   \Theta ^{i }{}_{j}=-P \delta ^{i }{}_{j }.
\eeq

Substituting the above perturbed variables into the field equation (\ref{EoM}),  we can get the background equations and linear perturbed equation via counting the orders of perturbations. With some cumbersome algebra, the background equations read
\begin{align}
&\frac{\sqrt{\left(1-B H^2\right)} }{\sqrt{1-A H^2}}\left(1+2 B H^2-3 A B H^4\right)-1=\frac{16 \pi  G}{\lambda }\rho, \label{BG_EoM1}\\
&\frac{\sqrt{\lt(1-B H^2\rt)^{-1}}}{\sqrt{\lt(1-A H^2\rt)^3}} \Bigg[1+\frac{A+3 B}{d} \dot{H}-(A- B) H^2\nn\\
&-\frac{14 AB+6 B^2}{3} H^2 \dot{H}-\left(4A B+ 2B^2\right) H^4\nn\\
&+\frac{\left(9 A+19B\right) A B}{3}H^4 \dot{H}+(3 A+5 B) A B H^6\nn\\
&-4 A^2 B^2 H^6 \dot{H}-3 A^2 B^2 H^8\Bigg]-1=-\frac{16 \pi  G}{\lambda } P,
\label{BG_EoM2}
\end{align}
where the constants $A=6(\beta+2\gamma)\lambda^{-1}$ and $B=2(2\alpha+\beta+6\gamma)\lambda^{-1}$.

Further, by counting the first-order perturbations of  field equation (\ref{EoM}), the only non-vanishing equation reads
\beq
F_2 \ddot{h}^{i}{}_j+F_1 \dot{h}^i{}_j-F_0 \nabla^2 h^i{}_j=0,
\label{Evolution_Eq}
\eeq
where the Laplace operator $ \nabla^2=\delta^{ij}\pt_i \pt_j$, and the coefficients are given by
\begin{align}
&F_0=\frac{4}{3 a^2} \left[\frac{D}{1-A H^2}+\frac{3-D\lambda}{\lambda  \left(1-B H^2\right)}\right],\\
&F_1=\frac{A H}{1-A H^2}+\frac{4 (B-D) H}{1-B H^2}+\frac{A^2 H \dot{H}}{3 \left(1-A H^2\right)^2}\nn\\
&-\frac{4 C H+2 \left(B^2+2 C^2\right) H \dot{H}-(3 B-4 C+4 D) C H^3}{2 \left(1-B H^2\right)^2}\nn\\
&-\frac{2 A B H \dot{H}}{\left(1-A H^2\right) \left(1-B H^2\right)}+\frac{ A C^2 H^3 \dot{H}}{ \left(1-A H^2\right) \left(1-B H^2\right)^2}\nn\\
&-\frac{B C^2 H^3 \dot{H}}{\left(1-B H^2\right)^3},\\
&F_2=\frac{A}{3 \left(1-A H^2\right)}+\frac{B}{1-B H^2}-\frac{C^2 H^2}{\left(1-B H^2\right)^2},
\end{align}
with the constants $C=(2 \alpha +\beta)\lambda^{-1} $ and $D=3\gamma\lambda^{-1}$. 

In low energy regime ($\lambda\rightarrow\infty$), the evolution equation (\ref{Evolution_Eq}) of the TT tensor mode reduces to the standard one in general relativity as expected, i.e.,
\beq
\ddot{h}^{i}{}_j+3 H \dot{h}^i{}_j+\frac{k^2}{a^2} h^i{}_j=0.
\eeq
where we have replaced the Laplace operator with $-k^2$ with $k$ the wave vector.

From the evolution equation (\ref{Evolution_Eq}), one can see that the perturbation dynamics of tensor mode are influenced by the coefficients $F_0$, $F_1$, and $F_2$, where $F_2$ is relevant to the kinetic term of the graviton and a negative-signed $F_2$ causes a ghost instability, and a negative-signed of $F_0$ is related to the gradient instability. For convenience, we recast  (\ref{Evolution_Eq}) as 
\beq
\ddot{h}^{i}{}_j+\mathcal{F} \dot{h}^i{}_j+\mathcal{M}^2 h^i{}_j=0,
\eeq 
where $\mathcal{F}={F_1}/{F_2 }$ is the effective friction coefficient, and $\mathcal{M}^2 =k^2{F_0} /{F_2 } $ is the effective squared mass. As time grows, a positive $\mathcal{F}$ generates an effective frictional force, which helps the system to be stabilized, but a negative $\mathcal{F}$ generates an effective accelerating force, which may lead to an $\mathcal{F}$-accelerating instability. However, if we are interested in the tensor evolution in the remote past, their roles are switched, where a positive $\mathcal{F}$ will generate an effective accelerating force but a negative $\mathcal{F}$ will generate frictional force. Moreover, if the effective squared mass $\mathcal{M}^2$ is negative, there may exist undesirable exponential growth. However, whether the system is unstable or not, it also depends on which one of  $\mathcal{F}$ and  $\mathcal{M}^2$ is dominant in the evolution \cite{Fukushima2019}. Here, some possible stabilities and instabilities of tensor evolution are classified as follows:

\textit{a})  If $F_2<0$, the theory suffers a ghost instability, which leads to a fatal collapse of vacuum in perturbations.

\textit{b}) If $F_2>0$, $F_0<0$, and $F_1<0$,  the coefficients $\mathcal{M}^2<0$ and $\mathcal{F}<0$. So generally speaking, the theory suffers a gradient instability. However, since a negative $\mathcal{F}$ generates an effective frictional force as time goes backwards, if this effect is dominant, the tensor evolution may be stabilized in the remote past.

\textit{c}) If $F_2>0$, $F_0<0$, and $F_1>0$,  the coefficients $\mathcal{M}^2<0$ and $\mathcal{F}>0$. In general, the theory suffers a gradient instability too. However, if the effective frictional force is dominant, an effective frictional force may stabilize the system in the remote future.

\textit{d}) If $F_2>0$, $F_0>0$, and $F_1<0$,  the coefficients $\mathcal{M}^2>0$ and $\mathcal{F}<0$. If $\mathcal{M}^2$ is dominant, the tensor evolution is stable. However, if $\mathcal{F}$ is dominant, the tensor evolution is unstable in remote future but stable in the remote past.

\textit{e}) If $F_2>0$, $F_0>0$, and $F_1>0$,  the coefficients $\mathcal{M}^2>0$ and $\mathcal{F}>0$. The tensor evolution is stable as time grows. As time goes backwards, if $\mathcal{M}^2$ is dominant, the tensor evolution is stable, however, if the effective accelerating force is dominant, the tensor mode may blow up in the remote past.

Furthermore, depending on the specific solutions, such as there are some poles in coefficients $\mathcal{F}$ and $\mathcal{M}^2$ at $t=0$, some other instabilities may involve in the tensor evolutions.  

In particular, if we consider the case of vacuum, i.e., $\rho=0$ and $P=0$, the background equations (\ref{BG_EoM1}) and (\ref{BG_EoM2}) lead to a constant Hubble parameter. Then the effective friction coefficient $\mathcal{F}$ and the effective squared mass $\mathcal{M}^2$ reduce to
\beqn
\mathcal{F} \!&\!&\!=\!3H,\\
\mathcal{M}^2 \!&\!&\!= \!\frac{4 }{a^2}\lt(3+\alpha -\beta -c\rt) \lt(3-2\alpha+2\beta-\alpha^2+\beta^2 -c\rt)\nn\\
\!\!&\!\!&\!
\Big[(1-\alpha +\beta ) \left(36+24\alpha-24\beta +16 \alpha ^2+4 \alpha \beta+7 \beta^2\right)\nn\\
\!\!&\!\!&\!- \lt( 24-4 \alpha+4\beta+12\alpha\beta-3\beta^2\rt)c+c^2\Big]^{-1},
\eeqn
where the constant $c={\lambda }/{H^2}$. So it allows us to check whether a specific vacuum solution is free of instabilities. Such as for the case of $\beta =\alpha +3$ (i.e., $B=0$), the solution is $H=0$, then $\mathcal{F} = 0$ and $\mathcal{M}^2=1/a^2$, the vacuum solution is free of instabilities. For the case of $\alpha=\beta$ (i.e., $A=B$), the solution is $H=0$ or $H=\sqrt{2\lambda/9}$. For $H=\sqrt{2\lambda/9}$, one has $\mathcal{F} = \sqrt{2\lambda}$ and $\mathcal{M}^2=2/a^2(2-3\alpha^2)$, the stability holds for the parameter $|\alpha|<\sqrt{2/3}$.

\section{Tensor evolutions of two specific solutions}\label{Sec_Stability}

It is reported that there are some cosmological solutions which can replace the Big Bang singularity with a de-Sitter phase or a bounce in the Born-Infeld determinantal gravity \cite{Fiorini2013,Fiorini2016}. Thus it is essential to investigate the tensor evolutions around the de-Sitter phase or the bounce point. Only if the background solution does not collapse under small perturbations, the assertions of singularity avoidance could be credible.

\subsection{Solution I}

An interesting solution discussed in Ref.~\cite{Fiorini2013} is achieved by choosing $B=0$. Combining the normalization condition $\alpha+\beta+4\gamma=1$, one has $\beta =\alpha +3$ and $\gamma = - (1+\alpha)/2 $ with $\alpha$ a free parameter. It leads to $A=12\lambda^{-1}$.  In this case, the background equations (\ref{BG_EoM1}) and (\ref{BG_EoM2}) reduce to
\beqn
\lt(1-\frac{12 H^2}{\lambda }\rt)^{-\frac{1}{2}}-1&=&\frac{16 \pi G}{\lambda }\rho,\\
\frac{ 1-\frac{16 H^2}{\lambda }-\frac{4 q H^2}{\lambda}}{\lt(1-\frac{12 H^2}{\lambda }\rt)^{\frac{3}{2}}}-1&=&-\frac{16  \pi G }{\lambda } P,
\eeqn
where $q=-\frac{a''}{a H^2}$ is the deceleration parameter. The square roots in the equations restrict the parameter $\lambda$ to be positive in this case.

By considering a perfect fluid with the state equation $P=\omega\rho$,
one can find that for every barotropic index $\omega>-1$,  the solution describes a geodesically complete spacetime without the big bang singularity and possesses a geometrical de Sitter inflationary stage naturally \cite{Fiorini2013,Ferraro2007,Ferraro2008}.
From the conservation equation $\dot \rho+3(\rho+P)H=0$, one has
\beq
\rho (t)=\rho_0 \left(\frac{a_0}{a(t)}\right)^{3 (\omega +1)},
\eeq
with the constants $a_0$ and $\rho_0$ relating to the present day values.

In the asymptotic past $t\rightarrow-\infty$, the Hubble factor approaches a maximum value $H_{\text{max}}\approx\sqrt{\lambda/12}$, and the scale factor behaves like
\beq
a(t\rightarrow-\infty)\propto e^{\sqrt{\frac{\lambda}{12}}\, t }\lt(1-\varepsilon e^{\sqrt{3\lambda}\lt (1+\omega\rt)\, t}\rt),
\eeq
where  $\varepsilon=\frac{1}{12 (\omega +1) }\left(\frac{16 \pi  G \rho_0}{\lambda }\right)^{-2}>0$.
Then it is easy to show that $\frac{\dot{H}}{1-A H^2}\approx- \lambda  (1+\omega)/4$, and
\beqn
F_2&\approx&\frac{e^{-\sqrt{3\lambda } (1+\omega ) t } }{3\varepsilon\lambda(1+\omega)}+\mathcal{O}(1), \label{Sol_I_1}\\
\frac{F_1}{F_2}&\approx&-\frac{\sqrt{3\lambda}}{2}  \omega+\mathcal{O}(e^{\sqrt{3\lambda } (1+\omega ) t } ),\label{Sol_I_2}\\
\frac{F_0}{F_2}&\approx&-\frac{e^{- \sqrt{\frac{\lambda}{3}} t}}{8} \left(4 (1+\alpha )-\varepsilon K_0 e^{\sqrt{3\lambda } (1+\omega ) t } \right),\label{Sol_I_3}
\eeqn
where $K_0=127+135 \omega +\alpha  [13+21 \omega -9 \alpha  (3+\alpha ) (1+\omega )]$.

\textit{a}) If $\alpha\neq-1$, we have $F_0/F_2\approx-\frac{1+\alpha }{2} e^{-\sqrt{\frac{\lambda }{3} } t}$.
The evolution equation is approximately to be
\beq
\ddot{h}^{i}{}_j-\frac{\sqrt{3\lambda}}{2}  \omega\dot{h}^i{}_j-\frac{(1+\alpha)k^2  }{2}  e^{-\sqrt{\frac{\lambda }{3} } t} h^i{}_j=0.
\eeq
For $\alpha<-1$, the solution is given by
\beqn
{h}^{i}{}_j\propto e^{\frac{\sqrt{3\lambda}}{4}\omega t}\lt[c_1\text{J}_\frac{3\omega}{2}\lt(p_1 k \sqrt{\frac{6}{\lambda}}e^{-\sqrt{\frac{\lambda}{12}}t} \rt)\rt.\nn\\
\lt.+c_2\text{J}_{-\frac{3\omega}{2}}\lt(p_1 k\sqrt{\frac{6}{\lambda}}e^{-\sqrt{\frac{\lambda}{12}}t} \rt)\rt],
\eeqn
where $\text{J}_n(z)$ is the Bessel function of the first kind, $p_1=\sqrt{-(1+a)}$, and $c_i$ are the integration constants. In the limit where $t\rightarrow-\infty$, the solution behaves like
\beqn
{h}^{i}{}_j\propto  e^{\sqrt{\frac{\lambda }{48}} (1+3 \omega)t }\lt[c'_1\cos\lt(p_1 k e^{-\sqrt{\frac{\lambda}{12}}t}-\frac{1+3\omega}{4}\pi \rt)\rt. \nn\\
\lt.c'_2\sin\lt(p_1 k e^{-\sqrt{\frac{\lambda}{12}}t}+\frac{1+3\omega}{4} \pi\rt)\rt].
\eeqn
When the barotropic index $\omega\geq-1/3$, the evolution of the tensor mode is non-divergent, hence, it holds stability in this case. However,  when $\omega<-1/3$, the tensor mode blows up as in the asymptotic past, so it renders an instability.

For $\alpha>-1$, the solution reads
\beqn
{h}^{i}{}_j\propto e^{\frac{\sqrt{3\lambda}}{4}\omega t}\lt[c_1 \text{I}_\frac{3\omega}{2}\lt(p_2 k \sqrt{\frac{6}{\lambda}}e^{-\sqrt{\frac{\lambda}{12}}t} \rt) \rt. \nn\\
\lt. +c_2 \text{I}_{-\frac{3\omega}{2}}\lt(p_2 k \sqrt{\frac{6}{\lambda}}e^{-\sqrt{\frac{\lambda}{12}}t} \rt)\rt],
\eeqn
where $\text{I}_n(z)$ is the modified Bessel function of the first kind and $p_2=\sqrt{1+a}$. In the limit $t\rightarrow-\infty$, the two Bessel functions are both proportional to $\exp\Big[\sqrt{{\lambda}/{48}}~t+p_2 k \sqrt{6/\lambda}~\exp\lt(-\sqrt{\lambda/12}~t\rt)\Big]$, so the solution behaves like
\beqn
{h}^{i}{}_j \propto \exp\lt[\sqrt{\frac{\lambda }{48}} (1+3 \omega)t +p_2 k \sqrt{\frac{6}{\lambda}}e^{-\sqrt{\frac{\lambda}{12}}t}\rt].
\eeqn
It will be divergent as $t\rightarrow-\infty$. Thus, the solution is unstable against the tensor perturbation in this case.

\textit{b}) If $\alpha=-1$ and $-1<\omega<-2/3$, we know that $F_0/F_2\approx 12 \varepsilon (1+\omega) e^{{\sqrt{\frac{\lambda }{3}} (2+3 \omega )t }}$ is divergent as $t\rightarrow-\infty$. The evolution equation  (\ref{Evolution_Eq}) reduces to
\beqn
\ddot{h}^{i}{}_j-\frac{\sqrt{3\lambda}}{2}  \omega\dot{h}^i{}_j+p_4^2 k^2 e^{-p_3 t}  h^i{}_j=0,
\eeqn
where $p_3=-\sqrt{\frac{\lambda}{3}}(2+3\omega)>0$ and $p_4=\sqrt{12 \varepsilon  (1+\omega)}$.
Then, the solution is given by
\beqn
{h}^{i}{}_j \propto e^{\frac{\sqrt{3 \lambda}}{4}\omega t}\lt[c_1\text{J}_\frac{3 \omega }{4+6 \omega } \lt(\frac{2 p_4 k}{p_3}e^{-\frac{p_3}{2}t} \rt)\rt.\nn\\
\lt.+c_2\text{J}_{-\frac{3 \omega }{4+6 \omega}}\lt(\frac{2 p_4 k}{p_3}e^{-\frac{p_3}{2}t} \rt)  \rt].
\eeqn
As $t\rightarrow-\infty$, the solution behaves like
\beqn
{h}^{i}{}_j \propto e^{-\sqrt{\frac{\lambda}{12}}t}\lt[c'_1\sin\lt(\frac{2p_4 k}{p_3}e^{-\frac{p_3}{2}t}+\frac{p_3+\sqrt{3\lambda}\omega }{4 p_3}\pi  \rt)\rt. \nn\\
\lt.+c'_2\cos\lt(-\frac{2p_4 k}{p_3}e^{-\frac{p_3}{2}t}+\frac{p_3+\sqrt{3\lambda}\omega }{4 p_3}\pi  \rt) \rt].
\eeqn
It will blow up in  the asymptotic past, thus the solution is unstable in this case.

\textit{c}) If $\alpha=-1$ and $\omega=-2/3$, we have $F_0/F_2\approx 4 \varepsilon$. Then, the evolution equation (\ref{Evolution_Eq}) is rewritten as
\beq
\ddot{h}^{i}{}_j-{\sqrt{\lambda/3}} \dot{h}^i{}_j+4 \varepsilon k^2 h^i{}_j=0.
\eeq
The solution is given by
\beq
h^i{}_j\approx c_1 e^{-\sqrt{\frac{\lambda }{12}} \lt(1+\sqrt{p_5}\rt) t}+c_2 e^{-\sqrt{\frac{\lambda }{12}}  \lt(1-\sqrt{p_5}\rt)t},
\eeq
where $p_5=1-\frac{48  \varepsilon}{\lambda }$. So the tensor mode blows up as $t\rightarrow-\infty$, and hence, the spacetime renders an instability within the tensor perturbation in this case.

\textit{d}) If $\alpha=-1, \omega>-2/3$ and $\omega\neq0$, $F_0/F_2\approx 12 \varepsilon (1+\omega) e^{{\sqrt{\frac{\lambda }{3}} (2+3 \omega )t }}$ is negligible comparing with $F_1/F_2$. So the asymptotic behavior of the evolution equation (\ref{Evolution_Eq}) is
\beq
\ddot{h}^{i}{}_j-\frac{\sqrt{3\lambda}}{2}  \omega\dot{h}^i{}_j=0,
\eeq
which can be solved as
\beq
h^{i}{}_j \approx\frac{2 c_1 }{\sqrt{3 \lambda } \omega }e^{\frac{1}{2} \sqrt{3 \lambda }\omega  t }+c_2.
\eeq
 As $t\rightarrow-\infty$, the tensor mode is convergent and stable for $\omega>0$, but divergent and unstable for $-2/3<\omega<0$.

\textit{e}) If $\alpha=-1$ and $\omega=0$, we have $F_1/F_2\propto e^{{\sqrt{{3\lambda }} t }}$ and $F_0/F_2\propto  e^{{\sqrt{\frac{4\lambda }{3}} t }}$, and we can simplify the evolution equation (\ref{Evolution_Eq}) as
\beq
\ddot{h}^{i}{}_j\approx 0.
\eeq
Thus the solution reads
\beq
{h}^{i}{}_j=c_1 t+c_2,
\eeq
and the tensor mode is linearly divergent in the asymptotic past.

In brief, the evolution of the tensor mode is regular in the very early cosmic stage for $\alpha<-1$, $\omega\geq-1/3$ and $\alpha=-1$, $\omega>0$. However, the tensor mode is unstable in the asymptotic past for other regions of the parameter space of $\alpha$ and $\omega$. From Eqs. (\ref{Sol_I_1}), (\ref{Sol_I_2}) and (\ref{Sol_I_3}), the coefficients $F_2>0$,  $\mathcal{M}^2=F_0/F_2>0$ and $\mathcal{F}=F_1/F_2<0$ for $\alpha= -1$ and $\omega>0$, so the theory gets rid of  ghost instability, gradient instability and $\mathcal{F}$-accelerating instability in this parameter space. For the parameter $\alpha < -1$, the positive $F_2$ and $\mathcal{M}^2$ ensure that the theory gets rid of  ghost and gradient instabilities. Then for $\omega \geq 0$, there is no $\mathcal{F}$-accelerating instability. For $-1/3\leq  \omega <0 $, a positve $\mathcal{F}$ generates an effective accelerating force as time goes backwards, however, the theory is still stable as $\mathcal{M}^2$ is dominant in this case. 

\subsection{Solution II}

Another interesting solution studied in Ref. \cite{Fiorini2016} is achieved by setting $A=B$, which implies $\alpha=\beta$ and leads to $A=B=3\lambda^{-1}$ with $\gamma$ a free parameter.
In this case, the background equations (\ref{BG_EoM1}) and (\ref{BG_EoM2}) reduce to
\beqn
3 H^2 \left(1-\frac{9 H^2}{2 \lambda }\right)&=&8 \pi  G \rho,\\
3 H^2 \left(1-\frac{9 H^2}{2 \lambda }\right)+2 \dot{H} \left(1-\frac{9 H^2}{\lambda }\right)&=&-8 \pi  G P.
\eeqn

For a perfect fluid with the Born-Infeld parameter $\lambda<0$, the solution depicts an irregular spacetime with the Hubble rate diverges as the scale factor goes to zero, whereas for $\lambda>0$ and  the barotropic index $\omega>-1$, a brusque bounce solution can be obtained with the approximate behavior around the bouncing point given by
\beqn
\frac{a(t)}{a_0}\!&\!\approx\!&\!\lt(\frac{48 \pi  G  \rho_0}{ \lambda}\rt)^{\frac{1}{3 (1+\omega )}}\!\!\lt[1\!\pm\frac{\sqrt{\lambda}}{3}t-\! \frac{\lambda^{\frac{3}{4}}(1+\omega)^{\frac{1}{2}}}{9}(\pm t)^{\frac{3}{2}}\rt.\nn\\
&&\lt.+\frac{\lambda(7-\omega)}{144}t^2 \rt]+\mathcal{O}(t^\frac{5}{2}),
\eeqn
where the positive and minus signs correspond to $t>0$ and $t<0$, respectively. The Hubble rate is given by
\beq
H(t)\approx\pm\frac{\sqrt{\lambda }}{3}\mp\frac{\lambda^{\frac{3}{4}}(1+\omega)^{\frac{1}{2}}}{6}(\pm t)^{\frac{1}{2}}-\frac{\lambda}{72}(1+\omega)t+\mathcal{O}(t^{\frac{3}{2}}).
\eeq
As $|t| \to 0$, the Hubble rate reaches a maximum $H(0)\approx\pm\sqrt{\lambda }/3$, where the maximum energy density is $\rho_{\text{m}}\approx(48\pi G)^{-1}\lambda$ and the scale factor reaches a minimum $a(0)\approx\left( \rho_0/\rho_\text{m}\right)^{\frac{1}{3 (1+\omega )}}$. Then the cosmic time derivative of the Hubble rate reads
\beq
\dot H(t)=-\frac{\lambda^{\frac{3}{4}}(1+\omega)^\frac{1}{2}}{12(\pm t)^\frac{1}{2}}-\frac{\lambda(1+\omega)}{72}+\mathcal{O}(t^{\frac{1}{2}}).
\eeq
Although the scale factor and the Hubble rate is finite, the cosmic time derivative of the Hubble rate diverges at the bouncing point. This divergence can also be directly seen from the Raychaudhuri equation \cite{Bouhmadi-Lopez2014a}
\beq
\dot H = -\frac{3(1+\omega)\rho}{2}\frac{dH^2}{d\rho},
\eeq
where $\frac{dH^2}{d\rho}=\frac{8 \pi  G}{3 \lt(1-\frac{9}{\lambda } H^2\rt)}$. It is clear that $\dot H$ diverges when the Hubble rate reaches its maximum $\sqrt{\lambda }/3$. Furthermore, because $t\propto \rho_\text{m}-\rho$, the divergence happens at a finite time and corresponds to a sudden singularity, which comes from a purely geometrical feature \footnote{The original solution in Ref. \cite{Fiorini2016} is claimed to be regular since the author only include the terms up to $\mathcal{O}(t)$ in the scale factor. However, by including the higher order terms, the solution exhibits a sudden singularity at the bounce.}.

With the asymptotic solution, the coefficients of the evolution equation (\ref{Evolution_Eq}) behave like
\beqn
F_0&\approx & \frac{6 }{a_0^2 \lambda }\left(\frac{\rho_\text{m}}{\rho_0}\right)^{\frac{2}{3 (1+\omega )}}+\mathcal{O}{(t^\frac{1}{2})},\\
F_1&\approx & \pm\frac{3K_1(1+\omega)^\frac{1}{2}}{32\lambda^\frac{3}{4} \sqrt{\pm t}}\pm\frac{144- (5+2 \omega )K_1}{16 \sqrt{\lambda }}\!+\!\mathcal{O}{(t^\frac{1}{2})},\\
F_2&\approx &-\frac{3 K_2}{16\lambda}+\frac{3(24+K_2) (1+\omega)^\frac{1}{2}}{8 \lambda ^{3/4}}{\sqrt{\pm t}}+\mathcal{O}{(t)},
\eeqn
where $K_1=48 \gamma ^2-24 \gamma +19$, $K_2= 48 \gamma ^2-24 \gamma -29$, and the positive and minus signs in $F_1$ and $F_2$ correspond to $t>0$ and $t<0$, respectively.

\textit{a}) Since $K_1$ is positive for all real $\gamma$, so if $K_2$ is non-null, i.e., $\gamma \neq \frac{1}{4}\pm\sqrt{\frac{2}{3}}$, $F_1/F_2\approx \mp\frac{\lambda^\frac{1}{4}(1+\omega)^\frac{1}{2}K_1}{2 K_2(\pm t)^\frac{1}{2}}$ and
$F_0/F_2\approx-\frac{32}{a_0^2 K_2}\left(\frac{\rho_{\text{m}}}{\rho_0 }\right)^{\frac{2}{3 (1+\omega )}} $. Then, the evolution equation (\ref{Evolution_Eq}) reads
\beq
\ddot{h}^{i}{}_j\pm\frac{q_1}{\sqrt{\pm t}}\dot h^i{}_j+q_2 k^2h^i{}_j=0,
\eeq
    where $q_1=-\frac{\lambda^\frac{1}{4}(1+\omega)^\frac{1}{2}K_1}{2 K_2}$, $q_2=-\frac{32 }{a_0^2 K_2}\left(\frac{\rho_{\text{m}}}{\rho_0 }\right)^{\frac{2}{3 (1+\omega )}}$, and the positive (minus) sign corresponds to $t>0$ ($t<0$). This equation is hard to be solved directly. So in order to explore the evolution behavior around the bouncing point analytically, we make a coordinate transformation $dt=e^{2 q_1\sqrt{\pm t(\tau)}}d\tau$, then the evolution equation can be rewritten as
\beq
\frac{d^2{h}^{i}{}_j}{d\tau^2}+q_2 k^2 e^{4q_1\sqrt{\pm t(\tau)}}h^i{}_j=0.
\label{Sol2a}
\eeq
From the coordinate transformation, we have
\beq
\tau=\pm\frac{1-e^{-2q_1\sqrt{\pm t}}(1+ 2q_1\sqrt{\pm t})}{2q_1^2},
\eeq
then $t\approx \tau \pm\frac{4}{3} q_1 (\pm\tau)^{3/2}+\mathcal{O}(\tau^2)$. Thus, around the bouncing point $|t|\to 0$ or $|\tau|\to 0$, the evolution equation (\ref{Sol2a}) can be approximated to be
\beq
\frac{d^2{h}^{i}{}_j}{d\tau^2}+q_2 k^2 h^i{}_j=0.
\eeq
The asymptotic behavior of the tensor mode around the bounce point is solved as
\beqn
h^i{}_j&\approx& c_1 \cos(\sqrt{q_2} k \tau)+c_2 \sin(\sqrt{q_2} k \tau),~(q_2>0)\\
h^i{}_j&\approx &c_1 e^{\sqrt{-q_2} k \tau}+c_2 e^{-\sqrt{-q_2} k \tau },~(q_2<0)
\eeqn
The tensor evolution is convergent around the bouncing point, therefore, the solution is stable under the tensor perturbation in this case.

\textit{b}) If $\gamma = \frac{1}{4}\pm\sqrt{\frac{2}{3}}$, then $K_1=48$ and $K_2=0$. Now, the asymptotic behavior of the evolution equation (\ref{Evolution_Eq}) reduces to
\beq
\ddot h^i{}_j+\frac{1}{2t}\dot h^i{}_j+\frac{q_3}{\sqrt{\pm t}} k^2h^i{}_j=0,
\eeq
where $q_3=\frac{2 }{3 a_0^2 \lambda^{1/4}\sqrt{(1+\omega)}}\left(\frac{\rho_{\text{m}}}{\rho_0 }\right)^{\frac{2}{3 (1+\omega )}}$ and the positive (minus) sign corresponds to $t>0$ ($t<0$).
Thus, the solution is given by
\beqn
{h}^{i}{}_j \propto (\pm t)^{\frac{1}{4}}\lt[c_1\text{BesselJ}\lt(\frac{1}{3},\frac{4\sqrt{q_3} k}{3}(\pm t)^{\frac{3}{4}} \rt)\rt.\nn\\
\lt.+c_2\text{BesselJ}\lt(-\frac{1}{3}, \frac{4\sqrt{q_3} k}{3} (\pm t)^{\frac{3}{4}} \rt)  \rt].
\eeqn
Its asymptotic behavior near the bouncing point is
\beq
{h}^{i}{}_j \propto c'_1 \sqrt{\pm t}+c'_2.
\label{Sol2casebpos}
\eeq
Therefore, the tensor evolution is convergent and stable near the bouncing point.

In brief, although the background solution suffers a sudden singularity, and there is a pole in $F_1$ at the bouncing point, the  evolution of the tensor mode is stable for all regions of the parameter space of $\gamma$ and $\omega$. 

\section{Parameter constraint from GW170817 and GRB170817A}\label{Sec_Parameter_Constraint}

The combined observation of  binary neutron star merger event GW170817 and its electromagnetic counterpart GRB170817A gives a strong constraint on the speed of gravitational waves, i.e., $-3\times 10^{-15}\leq (v_{GW}-c)/c \leq +7\times 10^{-15}$ \cite{Abbott2017}. 
In order to constrain the parameter of the theory, following the sprit of Ref.  \cite{Cai2018,Jana2017}, we consider the gravitational waves propagating in a background FRW universe. By utilizing the ansatz of the Fourier transformation as \cite{Cai2018}
\beq
{h^i}_{j}=\int{}d^3k e^{i\vec{k}\cdot\vec{x}}\lt[{A^i}_je^{i\omega t}+{B^i}_j e^{-i\omega t} \rt],
\label{Fourier_h}
\eeq
the evolution equation (\ref{Evolution_Eq}) reduces to 
\beq
\left(\omega ^2-\frac{F_0 }{F_2}k^2\right)\mp i \frac{F_1 }{F_2}\omega=0.
\label{Fourier_Evo_eq}
\eeq
 Then the dispersion relation is obtained as
 \beq
 \lt|\frac{d \omega }{dk}\rt|=\lt[\frac{F_2}{F_0}-\frac{F_1^2}{4F_0^2 k^2} \rt]^{-\frac{1}{2}}.
 \label{dispersion_relation}
 \eeq
 In our case, the wave lengths of gravitational waves are far shorter than the horizon, i.e., $k/a\ll H$. Moreover, the cosmological redshift for GRB170817A is $z\sim0.009$, thus the scale factor can be treated roughly as a constant during the propagation of gravitational waves. Now, in the limit of a small Hubble rate, the dispersion relation (\ref{dispersion_relation}) is approximated as
 \beq
  \lt|\frac{d \omega }{dk}\rt|\approx\frac{1}{a}\lt[1+\frac{9a^2H^2}{8k^2}+\frac{(2\alpha-\beta)(2\alpha+7\beta)H^2}{8\lambda}\rt].
 \eeq
 In order to get the common dispersion relation, we use a physical wave vector $\bar k=k/a$ and write the speed of light $c$ explicitly, then the above relation is rewritten as
  \beqn
  \lt|\frac{d \omega }{d\bar k}\rt|\approx c\lt[1+\frac{9H^2}{8 c^2 \bar k^2}+\frac{(2\alpha-\beta)(2\alpha+7\beta)H^2}{8 c^2 \lambda}\rt].
 \eeqn
 
 Since the gravitational waves propagating over an intergalactic distance in a FRW background, the second and third terms cause a deviation from the speed of light. In low energy regime ($\lambda\rightarrow\infty$), this equation reduces to the standard dispersion relation in general relativity. With the Hubble constant roughly $H_0\sim~70~\text{km}~\text{s}^{-1}~\text{Mpc}^{-1}$, it is easy to show that the second term is far less than $10^{-15}$, so this correction is negligible. Nevertheless, the third term comes form the effect of this modified gravity, hence, it is helpful to constrain the parameter of the theory with the bound on the speed of gravitational waves. Without loss of generality, we set the constants $\alpha, \beta$ to be of the order of 1 in the third term, then we can obtain the bound on the theoretical parameter $\lambda$ of the theory as $|\lambda|\geq 10^{-38} \text{m}^{-2}$.

\section{Conclusions}

We linearized the field equations of Born-Infeld determinantal gravity and investigated the tensor stability of two solutions in the early universe. For solution I, which supports a regular $\lambda$-driven de Sitter evolution of infinite duration, we found that the solution is stable against the tensor perturbation for $\alpha<-1$, $\omega\geq-1/3$ and $\alpha=-1$, $\omega>0$, yet the solution renders an instability for other parameter space. Therefore, for a very early radiation filled universe ($\omega=1/3$), about which we are most concerned, the cosmic evolution is stable against the tensor perturbation with the parameter $\alpha\leq-1$. For solution II, which supports a brusque bounce universe, we found that although the background solution exhibits a sudden singularity at the bouncing point, its tensor evolution is stable around the bounce for all regions of the parameter space of $\gamma$ and $\omega$. On the other hand, by taking into account the current bound on the speed of the gravitational waves, we obtained the bound on the theoretical parameter $\lambda$ as $|\lambda|\geq 10^{-38} \text{m}^{-2}$.

It is well-known that general relativity suffers the singularity problem in early universe, and the Born-Infeld type gravity may avoid the Big Bang singularity, such as the EiBI theory and the Born-Infeld determinantal gravity. However, although the background solution is singularity-free in EiBI theory, the overall evolution is singular because of the unstable tensor perturbation in Eddington regime. Therefore, our calculation suggests that the stable cosmic evolutions against tensor perturbation in large parameter spaces is a remarkable property of Born-Infeld determinantal gravity. The full linear perturbations, including scalar modes and vector modes, are quite more complicated and left for our future work.

\section*{ACKNOWLEDGMENTS}

The authors would like to thank B.-M. Gu and X.-L. Du for helpful discussions. This work was supported by the National Natural Science Foundation of China under Grant No. 11747021, No. 11875151, and No. 11522541. K. Yang acknowledges the support of ``Fundamental Research Funds for the Central Universities" under Grant No. XDJK2019C051. Y.-X. Liu acknowledges the support of ``Fundamental Research Funds for the Central Universities" under Grant No. lzujbky-2018-k11.

\appendix

\section{Perturbations of $\frac{\partial F_{\alpha\beta}}{ \partial {e^A}_\mu}$ and $\frac{\partial F_{\alpha\beta}}{\partial(\partial_\gamma {e^A}_\mu)}$}
\label{appendix}

The perturbation of $\frac{\partial F_{\alpha\beta}}{ \partial {e^A}_\mu}$ can be assembled by  $\frac{\partial F_{\alpha\beta}}{ \partial {e^A}_\mu}=\alpha \frac{\partial F^{(1)}_{\alpha\beta}}{ \partial {e^A}_\mu}
+\beta \frac{\partial F^{(2)}_{\alpha\beta}}{ \partial {e^A}_\mu}+\gamma \frac{\partial F^{(3)}_{\alpha\beta}}{ \partial {e^A}_\mu}$, with the nonvanishing components of $F^{(1)}_{\mu\nu}$, $F^{(2)}_{\mu\nu}$, and $F^{(3)}_{\mu\nu}$ given as
\begin{align}
\frac{\pt {F}^{(1)}_{ij}}{\pt e^0{}_0}&=-2(d-1) a^2 H^2\Big[  \delta _{ij}+2  h_{ij}+\frac{d-2}{d-1} \frac{ \dot{h}_{ij}}{H}\Big],\\
\frac{\pt {F}^{(1)}_{0 j}}{\pt {e}^0{}_k}&=-(d-1) H^2 \delta ^k{}_j-(d-2) H \dot{h}^k{}_j,\\
\frac{\pt {F}^{(1)}_{ij}}{\pt {e}^0{}_k}&=H\lt[{\pt_j h^k{}_i}-(d-2) {\pt_i h^k{}_j}+(d-3)\pt^k h_{ij}\rt],\\
\frac{\pt {F}^{(1)}_{i0}}{\pt{e}^a{}_0}&=\frac{\pt {F}^{(1)}_{0 i}}{\pt{e}^a{}_0}=(d-1)a H^2 \Big[ \delta _{ai}+ h_{ai}+\frac{d-2}{d-1}\frac{ \dot{h}_{ai}}{H}\Big],\\
\frac{\pt {F}^{(1)}_{ij}}{\pt {e}^a{}_0}&=a H \lt[\Big(d-\frac{3}{2}\Big) {\pt_i h_{aj}}-\frac{1}{2} {\pt_j h_{ai}}-(d-2) {\pt_a h_{ij}}\rt],\\
\frac{\pt{F}^{(1)}_{0 j}}{\pt{e}^a{}_k}&=\frac{H}{a}\lt(\frac{1}{2}\pt ^k h_{aj}+\frac{1}{2}{\pt_a h^k{}_j}-{\pt_j h_a{}^k}\rt),\\
\frac{\pt {F}^{(1)}_{ij}}{\pt{e}^a{}_k}&=a H^2 \lt[\Big(d-\frac{1}{2}\Big)  \delta ^k{}_j\delta _{ai}+\frac{1}{2}  \delta ^k{}_i\delta _{aj}-\delta _{ij}\delta _a{}^k \rt]\nn\\
&+a H^2 \Bigg[\Big(d-\frac{1}{2}\Big)  \delta ^k{}_jh_{ai}+\frac{1}{2} \delta ^k{}_i h_{aj} +\delta _{ij} h_a{}^k\nn\\
 &-2 \delta _a{}^k  h_{ij}\Bigg]+a H \Bigg[\frac{1}{2} \delta _{ai} \dot{h}^k{}_j+\delta _{aj} \dot{h}^k{}_i-\delta _{ij} \dot{h}_a{}^k\nn\\
 &-\delta _a{}^k\dot{h}_{ij}+\Big(d-\frac{3}{2}\Big) \delta ^k{}_j\dot{h}_{ai}  \Bigg],\\
\frac{\pt F^{(2)}_{ij}}{\pt {e}^0{}_0}&=-(d-1) a^2 H^2 \lt[ \delta _{ij}+2 h_{ij}+\frac{d-2}{d-1}\frac{ \dot{h}_{ij}}{H} \rt],\\
\frac{\pt F^{(2)}_{0 j}}{\pt {e}^0{}_k}&=-\frac{1}{2} \lt[(d-1) H^2 \delta ^k{}_j+(d-2) H \dot{h}^k{}_j\rt],\\
\frac{\pt F^{(2)}_{i0}}{\pt {e}^0{}_k}&=-(d-1) H^2 \delta ^k{}_i-(d-2) H \dot{h}^k{}_i,\\
\frac{\pt F^{(2)}_{ij}}{\pt e^0{}_k}&=\frac{H}{2}\lt[{\pt_i h^k{}_j}-(d-2) {\pt_j h^k{}_i}+(d-3)\pt ^kh_{ij}\rt],\\
\frac{\pt F^{(2)}_{i0}}{\pt e^a{}_0}&=\frac{\pt F^{(2)}_{0 i}}{\pt e^a{}_0}=\frac{d-1}{2}aH^2  \Big[  \delta _{ai}+ h_{ai}+\frac{d-2}{d-1}\frac{ \dot{h}_{ai}}{H}\Big],\\
\frac{\pt F^{(2)}_{ij}}{\pt e^a{}_0}&=\frac{a H}{2}  \lt[\Big(d-\frac{3}{2}\Big) {\pt_j h_{ai}}-\frac{1}{2} {\pt_i h_{aj}}-(d-2) {\pt_a h_{ij}}\rt],\\
\frac{\pt F^{(2)}_{00}}{\pt e^a{}_k}&=(d-1) \frac{H^2 }{a}\lt[{\delta _a}^k-  {h_a}^k+\frac{d-2}{d-1 } \frac{\dot{h}_a{}^k}{H} \rt],\\
\frac{\pt F^{(2)}_{i0}}{\pt {e^a}_k}&=\frac{H}{2a}\lt[(d-3) {\pt_i {h_a}^k}-(d-2) {\pt_a h^k{}_i}+\pt^k h_{ai}\rt],\\
\frac{\pt F^{(2)}_{0 j}}{\pt e^a{}_k}&=\frac{H}{2a}\lt[(d-2) {\pt_j h_a{}^k}-\lt(d-\frac{3}{2}\rt) {\pt_a h^k{}_j}+\frac{1}{2}\pt ^kh_{aj}\rt],\\
\frac{\pt F^{(2)}_{ij}}{\pt {e}^a{}_k}&=\frac{a H^2 }{4} \Big[\delta _{ai} \delta _j{}^k+(2 d-1) \delta _{aj} \delta_i{}^k-2 \delta _{ij} {\delta _a}^k\Big]\nn\\
&+\frac{a H^2}{4}  \Big[(2 d-1) {\delta _i}^k h_{aj}+{\delta_j}^k h_{ai} +
2 \delta _{ij} {h_a}^k\nn\\
&-4 {\delta _a}^k h_{ij} \Big]+\frac{a H}{4}  \Big[\delta _{ai} \dot{h}^k{}_j+2 (d-1){\delta_i}^k \dot{h}_{aj} \nn\\
&+{\delta_j}^k\dot{h}_{ai}-2 \delta _{ij} \dot{h}_a{}^k -2{\delta _a}^k \dot{h}_{ij} \Big],\\
\frac{\partial F^{(3)}_{ij}}{\pt {e^0}_0}&=-2 d (d-1)a^2  H^2 \lt(\delta _{ij }+2 h_{ij }\rt),\\
\frac{\partial F^{(3)}_{i0}}{\partial {e^0}_k}&=\frac{ \pt F^{(3)}_{0i} }{\pt {e^0}_k}=-d (d-1) H^2 \delta ^{k }{}_{i },\\
\frac{\partial F^{(3)}_{i0 }}{{\pt e}^a{}_0}&=\frac{\pt F^{(3)}_{0 i }}{{\pt e}^a{}_0}= d (d-1) a H^2\lt(\delta _{i a }+h_{i a }\rt),\\
\frac{\pt F^{(3)}_{00}}{ {\pt e}^a{}_{k}}&=2(d-1)\frac{H^2}{a} \lt[  \delta ^{k }{}_{a}- h^{k }{}_{a }+\frac{d-2}{d-1}\frac{\dot{h}^{k }{}_{a }}{H}\rt],\\
\frac{\pt F^{(3)}_{ij}}{\pt {e^a}_k}&=(d-1)a H^2 \lt(d \delta _{a(i}\delta ^k{}_{j)}-2 \delta ^k{}_a \delta _{ij}\rt)\nn\\
&+(d-1)a H^2 \left(d \delta ^k{}_{(i} h_{j)a}+2 \delta _{ij} h_a{}^k-4 \delta _a{}^k h_{ij} \rt)\nn\\
&-2(d-2) a  H \delta _{ij} \dot{h}_a{}^k.
\end{align}

The perturbation of $\frac{\partial F_{\alpha\beta}}{\partial(\partial_\gamma {e^A}_\mu)}$ can be assembled by
$\frac{\partial F_{\alpha\beta}}{\partial(\partial_\gamma {e^A}_\mu)}=\alpha \frac{\partial F^{(1)}_{\alpha\beta}}{\partial(\partial_\gamma {e^A}_\mu)}+\beta \frac{\partial F^{(2)}_{\alpha\beta}}{\partial(\partial_\gamma {e^A}_\mu)}+ \gamma \frac{\partial F^{(3)}_{\alpha\beta}}{\partial(\partial_\gamma {e^A}_\mu)}$, with the nonvanishing components of $\frac{\partial F^{(1)}_{\alpha\beta}}{\partial(\partial_\gamma {e^A}_\mu)}$, $\frac{\partial F^{(2)}_{\alpha\beta}}{\partial(\partial_\gamma {e^A}_\mu)}$, and $\frac{\partial F^{(3)}_{\alpha\beta}}{\partial(\partial_\gamma {e^A}_\mu)}$ given as
\begin{align}
\frac{\pt {F}^{(1)}_{i0}}{\pt (\pt_0 e^0{}_k)}&=-(d-1) H \delta ^k{}_i+\dot{h}^k{}_i,\\
\frac{\pt {F}^{(1)}_{ij}}{\pt  ({\pt_0 e^0{}_k)}}&=\partial ^kh_{ij}-{\pt_i h^k{}_j},\\
\frac{\pt F^{(1)}_{i0}}{\pt ({\pt_l e^0{}_0})}&=(d-1) H \delta ^l{}_i-\dot{h}^l{}_i,\\
\frac{\pt F^{(1)}_{ij}}{\pt ({\pt_l e^0{}_0})}&={\pt_i h^l{}_j}-\pt^l h_{ij},\\
\frac{\pt F^{(1)}_{0 j}}{\pt ({\pt_0 e^a{}_k})}&=\frac{1}{2a}\lt({\pt_a h^k{}_j}-\partial ^kh_{aj}\rt),\\
\frac{\pt F^{(1)}_{ij}}{\pt ({\pt_0 e^a{}_k})}&= a H \Bigg[\delta _a{}^k\delta _{ij}-\frac{1}{2} \delta _{ai} \delta ^k{}_j+\Big(d-\frac{3}{2}\Big) \delta _{aj} \delta ^k{}_i\Bigg]\nn\\
&-a \Bigg[\frac{H}{2} \delta _j{}^k h_{ai}-\Big(d-\frac{3}{2}\Big) H \delta ^k{}_i h_{aj}+H \delta _{ij} h_a{}^k\nn\\
&-2 H \delta _a{}^k h_{ij}+\delta _{aj } \dot{h}^k{}_i+\frac{1}{2} \delta _{ai} \dot{h}^k{}_j+\frac{1}{2} \delta ^k{}_i \dot{h}_{aj}\nn\\
&-\delta _a{}^k \dot{h}_{ij}\Bigg],\\
\frac{\pt F^{(1)}_{0 j}}{\pt (\pt_l e^a{}_0)}&=\frac{1}{2a}\lt(\partial ^l h_{aj}-{\pt_a h^l{}_j}\rt),\\
\frac{\pt F^{(1)}_{ij}}{\pt (\pt_l e^a{}_0)}&=-a H \Bigg[\delta _a{}^l \delta _{ij}-\frac{1}{2} \delta _{ai} \delta ^l{}_j+\Big(d-\frac{3}{2}\Big) \delta _{aj} \delta ^l{}_i\Bigg]\nn\\
&+a \Bigg[\frac{H}{2} \delta ^l{}_j h_{ai}-\Big(d-\frac{3}{2}\Big) H \delta ^l{}_i h_{aj}-2 H \delta _a{}^l h_{ij}\nn\\
&+H \delta _{ij} h_a{}^l+\delta _{aj} \dot{h}^l{}_i+\frac{1}{2} \delta _{ai} \dot{h}^l{}_j+\frac{1}{2} \delta _i{}^l \dot{h}_{aj}\nn\\
&-\delta _a{}^l \dot{h}_{ij}\Bigg],\\
\frac{\pt F^{(1)}_{i0}}{\pt (\pt_l e^0{}_k)}&=-{a^{-2}}\lt(\pt ^l h^k{}_i-\pt ^k h^l{}_i\rt),\\
\frac{\pt F^{(1)}_{0 j}}{\pt (\pt_l e^0{}_k)}&=-\frac{1}{2a^{2}}\lt(\pt ^l h^k{}_j-\pt ^kh^l{}_j\rt),\\
\frac{\pt F^{(1)}_{i j}}{\pt (\pt_l e^0{}_k)}&=\frac{1}{2} \lt(H \delta ^l{}_i \delta ^k{}_j-H \delta ^l{}_j \delta ^k{}_i+\delta ^l{}_i \dot{h}^k{}_j-\delta ^k{}_i \dot{h}^l{}_j\rt),\\
\frac{\pt F^{(1)}_{0j}}{\pt (\pt_l e^a{}_k)}&={a^{-1}}\Big(H \delta _a{}^k \delta ^l{}_j-H \delta _a{}^l \delta ^k{}_j+H \delta _j{}^k \delta _a{}^m h^l{}_m\nn\\
&-H \delta _j{}^l \delta _a{}^m h^k{}_m+\delta _a{}^k \dot{h}^l{}_j-\delta _a{}^l \dot{h}^k{}_j\Big),\\
\frac{\pt F^{(1)}_{ij}}{\pt (\pt_l e^a{}_k)}&={a^{-1}}\bigg[\frac{1}{2}\delta ^k{}_i\lt(\pt ^l h_{aj}-{\pt_a h^l{}_j}\rt)+\frac{1}{2}\delta ^l{}_i\big({\pt_a h^k{}_j}\nn\\
&-\pt ^k h_{aj}\big)+\frac{1}{2}\delta _{ai}\lt(\pt ^l h^k{}_j-\pt ^k h^l{}_j\rt)+\delta _{aj}\big(\pt ^lh^k{}_i\nn\\
&-\pt ^kh^l{}_i\big)+\delta _a{}^k\lt({\pt_i h^l{}_j}-\pt ^lh_{ij}\rt)+\delta _a{}^l\big(\pt ^k h_{ij}\nn\\
&-{\pt_i h^k{}_j}\big)\bigg],\\
\frac{\pt F^{(2)}_{i0}}{\pt (\pt_0 e^0{}_k)}&=\frac{1}{2} \lt[(d-1) H \delta^k {}_i-\dot{h}^k{}_i\rt],\\
\frac{\pt F^{(2)}_{ij}}{\pt (\pt_0 e^0{}_k)}&=\frac{1}{2}\lt(\pt ^k h_{ij}-{\pt_j h^k{}_i}\rt),\\
\frac{\pt F^{(2)}_{i0}}{\pt (\pt_l e^0{}_0)}&=-\frac{1}{2} \lt[(d-1) H \delta ^l{}_i-\dot{h}^l{}_i\rt],\\
\frac{\pt F^{(2)}_{ij}}{\pt (\pt_l e^0{}_0)}&=-\frac{1}{2}\lt(\pt ^l h_{ij}-{\pt_j h^l{}_i}\rt),\\
\frac{\pt F^{(2)}_{0 j}}{\pt (\pt_l e^0{}_k)}&=\frac{1}{4a^2}\left(\partial ^kh^l{}_j-\partial ^l h^k{}_j\right),\\
\frac{\pt F^{(2)}_{ij}}{\pt (\pt_l e^0{}_k)}&=\frac{1}{4} \lt(H \delta _i{}^k \delta ^l{}_j-H \delta _i{}^l \delta ^k{}_j+\delta _i{}^k \dot{h}^l{}_j-\delta _i{}^l \dot{h}^k{}_j\rt),\\
\frac{\pt F^{(2)}_{00}}{\pt (\pt_0 e^a{}_k)}&=-\frac{d-1}{a}\lt(H \delta _a{}^k-H h_a{}^k-\frac{\dot{h}_a{}^k}{d-1}\rt),\\
\frac{\pt F^{(2)}_{i0}}{\pt ({\pt_0 e^a{}_k})}&=\frac{1}{2a}\lt({\pt_i h_a{}^k}-\pt ^kh_{ai}\rt),\\
\frac{\pt {F}^{(2)}_{0 j}}{\pt  (\pt_0 e^a{}_k)}&=\frac{1}{4a}\left(2 {\pt_j h_a{}^k}-\partial ^k h_{aj}-{\pt_a h^k{}_j}\right),\\
\frac{\pt {F}^{(2)}_{i j}}{\pt  (\pt_0 e^a{}_k)}&=\frac{a}{2} \Bigg[H \delta _a{}^k \delta _{ij}+\Big(d-\frac{3}{2}\Big) H \delta _{ai} \delta ^k{}_j-\frac{H}{2}  \delta _{aj} \delta ^k{}_i\nn\\
&+\Big(d-\frac{3}{2}\Big) H \delta ^k{}_j h_{ai}-\frac{1}{2} H \delta^k{}_i h_{aj}-H \delta _{ij} h_a{}^k\nn\\
&+2 H \delta _a{}^k h_{ij}-\delta ^k{}_j \dot{h}_{ai}-\frac{1}{2} \delta^k{}_i \dot{h}_{aj}-\frac{1}{2} \delta _{ai} \dot{h}^k{}_j\nn\\
&+\delta _a{}^k \dot{h}_{ij}\Bigg],\\
\frac{\pt F^{(2)}_{00}}{\pt (\pt_l e^a{}_0)}&=\frac{d-1}{a}\lt[ H \delta _a{}^l-H h_a{}^l-\frac{\dot{h}_a{}^l}{d-1}\rt],\\
\frac{\pt F^{(2)}_{i0}}{\pt (\pt_l e^a{}_0)}&=-\frac{1}{2a}\lt({\pt_i h_a{}^l}-\partial ^l h_{ai}\rt),\\
\frac{\pt F^{(2)}_{0 j}}{\pt (\pt_l e^a{}_0)}&=-\frac{1}{4a} \lt(2 {\pt_j h_a{}^l}-\pt^l h_{aj}-{\pt_a h^l{}_j}\rt),\\
\frac{\pt F^{(2)}_{i j}}{\pt (\pt_l e^a{}_0)}&=-\frac{a}{2}  \Bigg[H \delta _a{}^l \delta _{ij}+\Big(d-\frac{3}{2}\Big) H \delta _{ai} \delta ^l{}_j-\frac{H}{2} \delta _{aj} \delta^l{}_i\nn\\
&-\frac{H}{2} \delta_i{} ^lh_{aj}+\lt(d-\frac{3}{2}\rt) H \delta _j{}^l h_{ai}-H \delta _{ij} h_a{}^l\nn\\
&+2 H \delta _a{}^l h_{ij}-\frac{1}{2} \delta_{ai} \dot{h}^l{}_j -\delta _j {}^l\dot{h}_{ai}-\frac{1}{2} \delta _i{}^l \dot{h}_{aj}\nn\\
&+\delta _a{}^l \dot{h}_{ij}\Bigg],\\
\frac{\pt F^{(2)}_{i j}}{\pt (\pt_l e^a{}_k)}&=\frac{1}{4a}\Big[2\delta _a{}^k \big({\pt_j h^{l }{}_i}-\pt ^{l }h_{ij}\big)-2\delta _a{}^{l} \big({\pt_j h^k{}_i}\nn\\
&-\pt ^k h_{ij}\big)+\delta _{ai}\big(\pt ^{l}h^k{}_j-\pt ^kh^{l }{}_j\big)-\delta _i{}^k\big(2 {\pt_j h_a{}^{l }}\nn\\
&-\pt ^{l }h_{aj}-{\pt_a h^{l }{}_j}\big)+\delta _i{}^{l }\big(2 {\pt_j h_a{}^k}-\pt ^k h_{aj}\nn\\
&-{\pt_a h^k{}_j}\big)-2 \delta ^k{}_j\big({\pt_i h_a{}^{l }}-\pt ^{l } h_{ai}\big)+2 \delta ^{l }{}_j\big({\pt_i h_a{}^k}\nn\\
&-\pt ^k h_{ai}\big)\Big],\\
\frac{\pt F^{(3)}_{00}}{\pt ({\pt_0 e^a{}_k})}&=\frac{2(1-d) }{a} \lt[H \delta _a{}^k- H h_a{}^k-\frac{\dot{h}_a{}^k}{d-1}\rt],\\
\frac{\pt F^{(3)}_{ij}}{\pt ({\pt_0 e^a{}_k})}&=2 a \delta _{ij} \lt[(d-1) H \delta_a{}^k-(d-1) H h_a{}^k-\dot{h}_a{}^k\rt]\nn\\
&+4 (d-1) a H \delta _a{}^k h_{ij},\\
\frac{\pt F^{(3)}_{00}}{\pt (\pt_l e^a{}_0)}&=  \frac{2(d-1)}{a}\lt[H \delta _a{}^l- H h_a{}^l-\frac{\dot{h}_a{}^l}{d-1}\rt],\\
\frac{\pt F^{(3)}_{ij}}{\pt ({\pt_l e^a{}_0})}&=-2 a \delta _{ij} \lt[(d-1) H \delta _a{}^l-(d-1) H h_a{}^l-\dot{h}_a{}^l\rt]\nn\\
&-4 (d-1) a H \delta _a{}^l h_{ij},\\
\frac{\pt F^{(3)}_{00}}{\pt ({\pt_l e^a{}_k})}&=-2{a^{-3}}\lt(\pt ^l h_a{}^k -\pt ^{k }h_a{}^l \rt),\\
\frac{\pt F^{(3)}_{ij}}{\pt ({\pt_l e^a{}_k})}&=2 {a^{-1}}\delta _{ij}\lt(\pt ^l h_a{}^k-\pt ^k h_a{}^l\rt).
\end{align}


\providecommand{\href}[2]{#2}\begingroup\raggedright\endgroup

\end{document}